%% file: main_arxiv.tex
\documentclass{article}
\usepackage[utf8]{inputenc}
\usepackage{graphicx}
\usepackage{authblk}
\usepackage{amssymb}
\usepackage{subfig}

\title{Testing cosmic ray composition models with very large volume neutrino telescopes}

\author[1]{L.A. Fusco}
\author[1,2,3]{F. Versari}
\affil[1]{\footnotesize{APC, AstroParticule et Cosmologie, Universit\`e Paris Diderot, CNRS/IN2P3, 10, rue Alice Domon et L\`eonie Duquet, 75205 Paris Cedex 13, France}}
\affil[2]{\footnotesize{Dipartimento di Fisica e Astronomia dell’Università, Viale Berti Pichat 6/2, I-40127 Bologna, Italy}}
\affil[3]{\footnotesize{INFN, Sezione di Bologna, Viale Berti-Pichat 6/2, I-40127 Bologna, Italy}}

\date{December 2019}

\begin{document}

\maketitle

\begin{abstract}
  
  The composition in terms of nuclear species of the primary cosmic ray flux is largely uncertain in the knee region and above, where only indirect measurements are available. The predicted fluxes of high-energy leptons from cosmic ray air showers are influenced by this uncertainty. Different models have been proposed. Similarly, these uncertainties affect the measurement of lepton fluxes in very large volume neutrino telescopes. Uncertainties in the cosmic ray interaction processes, mainly deriving from the limited amount of experimental data covering the particle physics at play, could also produce similar differences in the observable lepton fluxes and are affected as well by large uncertainties. In this paper we analyse how considering different models for the primary cosmic ray composition affects the expected rates in the current generation of very large volume neutrino telescopes (ANTARES and IceCube). We observe that, a certain degree of discrimination between composition fits can be already achieved with the current IceCube data sample, even though in a model-dependent way. The expected improvements in the energy reconstruction achievable with the next generation neutrino telescopes is be expected to make these instruments more sensitive to the differences between models.
  
\end{abstract}

\input{intro.tex}
\input{vlvnt.tex}
\input{atmocr.tex}
\input{atmonu.tex}
\input{stat.tex}

\input{results.tex}

\input{biblio.tex}
\end{document}

%% file: intro.tex
\section{Introduction} \label{intro}

When a primary cosmic ray (CR) reaches the top of the atmosphere, it penetrates it until it collides with an air nucleus. This usually happens at an altitude of about 10-20 km \cite{bib:gaisser}. An extensive cascade of particles, showering down in the atmosphere, results from this interaction. The most abundant hadronic products are pions and kaons, neutral or charged. Neutral pion decays induce an electromagnetic cascade in the atmosphere; charged mesons can decay leptonically. These leptons constitute the entirety of the atmospheric events that can be detected in a very large volume neutrino telescope (VLV$\nu$T), placed at large depths under sea, lake water, or ice. Even though these atmospheric leptons represent the foreground to the cosmic searches to which VLV$\nu$Ts are devoted, these experiments can provide an insight into the study of atmospheric lepton fluxes and thus into CR physics.

Extensive air-shower arrays indirectly measure the cosmic ray flux at Earth by observing the shower products arriving on the ground, coming from the electromagnetic component and the hadronic component -- the latter being addressed by measuring muons penetrating underground. The observed distributions of these particles can be correlated to the primary energy and species. Indeed, the direct measurements obtained by satellite and balloon flights, using magnetic deflection, allow accessing the mass spectrum of primary CR up to some tens of TeV; the per-species power-law behaviour could actually be extrapolated to higher energy but then the spectral features present above hundreds TeV/few PeV energies (namely the CR \textit{knee}) appear. A proper extrapolation is thus not bound by direct measurements since, because of the low flux, a proper measurement is not fully achievable. Indeed, in this energy range, CR air showers can be only observed by ground-based experiments, where in general only a distinction between a lighter and a heavier component could be obtained. The behaviour of each individual CR species around the knee and beyond is affected by large uncertainties, and current measurements do not completely agree in the description of the observed spectra. An overview of these factors, as well as of those measurement pointing at possible features in the CR spectrum will be presented in section \ref{sec:hecr} and their effect on the observable lepton fluxes in section \ref{sec:crnu}. 

The flux of atmospheric neutrinos can be measured, from sub-GeV to few hundreds-TeV energies, with the current generation of neutrino detectors and, in particular, with VLV$\nu$Ts, described in section \ref{sec:VLVnT}. This flux is the convolution of the primary cosmic ray flux at the top of the atmosphere and the neutrino yield per primary particle. This yield is affected by several factors, mostly related to the efficiency in producing the neutrino mother particles, \textit{i.e.} hadrons decaying into leptons. At a first approximation, the neutrino flux can be written as
\begin{equation}
  \frac{d\Phi_\nu} {dE_\nu d\Omega} (E_\nu,\theta) =  A E^{-\gamma_p} \sum_{m} \frac{ B_m }{ 1+\frac{c_m E_\nu}{\epsilon_m } \cos\theta}
  \end{equation}
where the first term $A E_\nu^{-\gamma_p}$ accounts for the primary CR spectrum, assumed \textit{e.g.} to behave as a power law with spectral index $-\gamma_p$. The sum then runs over the neutrino mother particles $m$: the coefficients B$_m$ and c$_m$ are related to the physics of the hadronic interaction at play. Finally, the term $\epsilon_m$ is the so-called \textit{critical energy}, for which the interaction length equals the decay length of the particle \cite{bib:gaisser}; above this energy all the mother particles would interact before decaying, thus reducing the quantity of neutrinos at the highest energies and making the spectrum steeper. 

For muon neutrinos below 100 TeV, $m = \{\pi,\, \kappa\}$, almost exclusively; this represent the \textit{conventional} atmospheric neutrino flux. At higher energy, shorter-lived particles such as charmed hadrons start contributing to the overall flux, producing the harder \textit{prompt} atmospheric neutrino component. More details will be given in section \ref{sec:crnu}.

Being the decay kinematics well-known, the overall flux is determined mostly by the primary flux, and by the production efficiency and interaction probability of the parent hadrons. Our knowledge of the former is affected by large uncertainties, both for what concerns the spectral behaviour of the individual components of the flux, and the nature itself of the primary particles species. The latter, \textit{i.e.} the description of the primary hadronic interaction and the description of the further interactions of these hadrons as the cascade of particles showers-down in the atmosphere, is rather uncertain because the interactions at play are low-transferred-momentum processes, not fully studied at accelerator experiments since the pseudo-rapidity ranges at play, are largely outside the reach for current detectors at large hadronic colliders, and similarly the parton distribution functions used in the description of these interactions are affected by large uncertainties, both in the theoretical models and in the experimental results.

In order to determine the effect of different CR composition models in the measurement of lepton fluxes at VLV$\nu$T, a simplified analysis is presented in this paper, taking into account the current generation of neutrino telescopes, built or currently being built. A Monte Carlo study of the possible performance of VLV$\nu$Ts in the study of atmospheric neutrino fluxes is set-up in section section \ref{sec:ingredients}. The statistical tools used in the analysis, including an overview of the systematic effects affecting the possible measurement are presented in section \ref{sec:stat}. The results of this study are shown in section \ref{sec:results}.

%% file: vlvnt.tex
\section{Very large volume neutrino telescopes} \label{sec:VLVnT}

High-energy neutrinos\footnote{Here and in the following, we refer to both $\nu$ and $\bar{\nu}$ as \textit{neutrinos}, unless differently specified} can be detected in a large volume of transparent medium (ice or water) by observing the Cherenkov light emitted by the relativistic particles induced by neutrino interactions. This light can be detected with a three-dimensional array of photo-multiplier tubes (PMTs), being these PMTs hosted in pressure-resistant glass spheres and distributed along a certain number of vertical strings to cover a large volume, large enough to allow the weakly-interacting-only neutrinos to produce a measurable amount of events in the detector.

Charged current (CC) weak interactions of muon neutrinos produce a long-lived relativistic muon, that can be identified as a straight track passing through the detector volume. Neutral current (NC) neutrino interactions, as well as CC interactions of electron neutrinos, would induce hadronic and electromagnetic particle showers in the medium, which can be identified as almost-point-like light sources, given the short elongation of these particle cascades with respect to the spacing between photo-sensors. Also tau neutrino interaction usually produce shower events, even though in CC interactions the tau lepton can travel a few tens of meters and be properly identified, allowing the observation of two separate particle showers.

Thanks to the long lever-arm of the passing-through track, track-like events provide sub-degree angular resolution, while the almost-spherical pattern of light in the detector given by cascades does not allow an extremely precise direction reconstruction (few degrees in water, roughly ten degrees in ice). On the other side, cascade-like events allow for an almost calorimetric measurement of the deposited energy, since most of the light is emitted within a limited amount of space. In contrast, the energy resolution in the case of track-like events is limited by the fact that only a part of the muon track is observed within the instrumented volume, and this muons usually arrives at the detector after having travelled large distances in the medium and thus after having lost a significant amount of its original energy. 

The measurement of the muon energy is related to the energy loss processes at play as the muon travels through the detector. Light is emitted along the track, and above 500 GeV in water/ice the muon loses most of its energy because of radiative processes, which are proportional to its energy. Radiative losses induce cascades of charged particles along the track, whose light yield is proportional to the energy lost in the event. Thus, a measurement of the energy loss along the track \textit{dE/dX} provides an estimate of the muon energy.

Track-like events in VLV$\nu$Ts allow for a selection of a  very pure sample of muon neutrino-induced events; in addition, the long path travelled by the muon increases the volume into which the neutrino interaction can take place, thus significantly increasing the number of detectable events. The larger statistics accessible with this kind of events can be extremely valuable for the study of atmospheric leptons. Event samples made of cascades present more limitations in terms of fiducial volume and background contamination.

\subsection{Currently active VLV$\nu$Ts}

Four large volume neutrino telescopes are currently taking data: IceCube at the geographic South Pole and ANTARES in the Mediterranean Sea have been completed about 10 years ago. Baikal-GVD, in Lake Baikal, Russia, and the KM3NeT/ARCA telescope in the Mediterranean Sea, off-shore the coasts of Sicily, are currently being build.

The IceCube detector is composed of 86 in-ice, 1-km-long, vertical strings, each holding 50 10-inch PMTs, covering a volume of 1 km$^3$ at a depth between 1500 and 2500 m in the Antarctic Ice Shell. The IceCube collaboration has provided the first observations of a high-energy cosmic neutrino signal, using different data samples \cite{bib:ic_hese}\cite{bib:ic_numu}\cite{bib:ic_casc} and of a first candidate cosmic source of high energy neutrinos \cite{bib:txs}.

The ANTARES telescope is made of 12 vertical lines, each holding 75 10-inch PMTs, instrumenting $\sim$0.01 km$^3$  of deep-sea waters off-shore the coasts of Southern France, at a depth of 2000-2500 m \cite{bib:ant}. Having been taking data continuously since 2007, ANTARES provides the longest neutrino telescope data-stream in the Northern Hemisphere. The currently available data-set has already let to significant results in searches for neutrino emissions from the Southern sky \cite{bib:ant_ps}\cite{bib:ant_gp} as well as the observation of a mild excess of high-energy neutrino which could be attributed to a diffuse cosmic neutrino flux \cite{bib:ant_dif}. 

The Baikal-GVD experiment, currently being deployed in Lake Baikal, has started data taking in 2015 and is at the moment the largest-volume neutrino detector in the Northern hemisphere \cite{bib:bai}. Analyses of shower-like events have already shown results in searches for cosmic neutrino fluxes \cite{bib:bai_dif}. At the moment no results are available for muon neutrinos, thus the possible performance of the Baikal-GVD detector is not studied in this paper.

The KM3NeT Collaboration aims at building a km$^3$-scale neutrino detector (ARCA) off the coasts of Sicily, at a depth of 2800-3500 m \cite{bib:km3_loi}. The first string, with its 18 Digital Optical Modules, made of 31 3-inch PMTs, is currently taking data. The goal of KM3NeT/ARCA is to do all-neutrino-flavour astronomy, and discover galactic neutrino emitters. Along with this, a highly significant observation of an IceCube-like diffuse signal is expected within a short time with the full detector, thus allowing for an independent confirmation of this signal. KM3NeT/ARCA is expected to outperform the current generation of neutrino telescopes in terms of angular and energy resolution as well as in the purity of the neutrino sample \cite{bib:jgandalf}.

%% file: atmocr.tex
\section{High-energy cosmic rays at Earth} \label{sec:hecr}

The per-particle spectra of primary cosmic rays can be measured with rather good precision using balloon-borne experiments, as well as with experiments on satellite in low Earth orbit. These measurements span energies from the sub-GeV range to several tens of TeV. Elements heavier than iron can also be identified and their spectrum can be measured. A review of recent measurements can be found in reference \cite{bib:cr_review}. As a matter of fact, the per-particle spectra follow power-law behaviour over the full range of energies covered by direct CR measurement, except for small kinks whose significance is not too relevant here.

Due to load limitations for satellites and balloons, these experiments cannot be too large and thus fail at measuring with the same precision the spectra of particles at very high energy/rigidity.  For primary energies above a hundred of TeV, only indirect measurement of the CR flux are possible. These measurements are based on the observation of the cascade of particles induced by the CR interaction at the top of the atmosphere. Different technologies are exploited to measure the properties of the cascade of particles, either as it develops through the atmosphere, or as it arrives on the ground, or when these particles reach a certain depth -- or, also, by means of a combination of these measurements. Indirect measurements go from 100 TeV to 100 EeV, but fail at identifing the nature of the CR nuclei on an event-by-event basis. 

It is however possible to obtain, by means of statistical methods and using Monte Carlo simulations, an estimation of the CR composition at the highest energy as either light (\textit{i.e} dominated by protons and helium nuclei) or heavier (whatever higher in atomic number). In any case these measurements are largely dependent on the CR interaction model as well as on the systematic effects in the estimation of the CR properties. The recent results from the Pierre Auger Observatory hint at a CR composition becoming heavier at the highest energies \cite{bib:pao}, even though some tension is present with the most recent results obtained by the Telescope Array Collaboration which prefer a lighter composition at the highest energies \cite{bib:ta}, above 10$^{19}$ eV.

At intermediate energies, around the knee region, several experiments have attempted to measure the composition on the basis of the shower topology and to perform per-particle spectral measurements; even though these measurements are as strongly model dependent on the hadronic interaction model considered, this has provided some hints on the behaviour of these components. In particular the KASCADE-Grande \cite{bib:kascade} and the ARGO-YBJ \cite{bib:argo} have tested the composition of the CR spectrum around the knee and actually do find different spectral breaks for the the light component. It should be noted that these measurements are in general affected by energy scale uncertainties which could end-up mimicking such behaviours. The LHAASO experiment \cite{bib:lhaaso} is expected to provide high-precision measurement of the CR spectrum in the knee region \cite{bib:lhaaso_cr} while Auger-Prime \cite{bib:auger_prime} should improve the knowledge of the very-high energy spectrum of CRs.

Various attempts have been made to reconstruct the all-particle flux in terms of its individual components, \textit{e.g} assuming power-law shape extrapolation from direct composition and flux measurements combined with rigidity-dependent cut-offs to accomodate the behaviour in the knee region and beyond. An overview of these possible fits to the all-particle spectrum is provided in reference \cite{bib:gst}. In general, several mass groups are considered for the primary composition and different fits become possible to describe the spectrum. In addition, different source populations can be assumed, inducing different rigidity-dependent cutoffs. Some example are the so-called \textit{Hillas models} (H3a and H4a, in the following, assuming 3 or 4 source populations) or the \textit{Gaisser-Stanev-Tilav models} (GST-3gen or GST-4gen in the following). A more global approach is followed by the authors of the Global Spline Fit \cite{bib:splinefit}, where the constraints on the power-law spectral behaviour and rigidity dependent cut-offs are relaxed.

In general, given the same per-nucleon energy of the primary CR particle, heavier nuclei would generate more individual showers due to the fragmentation of the nucleus as it traverse the atmosphere. This would change the content of the overall air shower at the bottom of the atmosphere and subsequentely the amount of muons and neutrinos reaching large depth. The next section will show the results of neutrino flux simulations performed with different CR composition fit results.

%% file: atmonu.tex
\section{Neutrinos in cosmic ray air showers} \label{sec:crnu}

In order to get the neutrino flux induced by cosmic ray air showers at the surface of the Earth, thus before they can be detected in a VLV$\nu$T, this flux must be either computed analytically or obtained from a simulation of the CR shower. The standard framework used for a full simulation of the CR air shower is \textit{CORSIKA} \cite{bib:corsika}. This framework allows for a detailed simulation of air shower with different possible assumptions, and the full propagation of particles through the atmosphere accounting for energy loss processes, as well as interactions. In addition, also the atmospheric profile can be customised, thus its influence on the final lepton fluxes can be taken into account.

Analytical solutions are possible for the computation of lepton fluxes through the atmosphere \cite{bib:gaisser}, even though these can only be considered approximations since some assumptions have to be made in order to make the calculation doable. The precision of these calculation is discussed in reference \cite{bib:gaisser_precision}.

The \textit{MCEq} software \cite{bib:mceq} has been developed to perform a precise computation of the evolution of particle densities along the cascade in the Earth's atmosphere. The cascade equations are solved by means of matrix computations taking into account the particle physics of user-selected hadronic interaction models such as \textit{e.g. SIBYLL 2.3c} \cite{bib:sibyll} or \textit{EPOS-LHC} \cite{bib:epos}, as well as cosmic ray input fluxes \cite{bib:crflux}. Similarly, user-defined atmosphere density profiles can be used, as in the CORSIKA framework. These features will be used extensively here to benchmark the performances of VLV$\nu$T in measuring features of the neutrino spectrum at the detector.

Above 100 GeV and below a few hundreds of TeV the flux of atmospheric muon neutrinos is dominated by the so-called conventional component, induced by the decays of long-lived charged mesons (basically pions and kaons). In the 10-100 TeV energy range, the muon neutrino flux is almost entirely produced by kaon decays \cite{bib:anatoli_sibincl}. In this region of the energy spectrum, the conventional flux is asymptotically steeper than the primary CR spectrum by a factor $\propto E^{-1}$ because of the competition between decay and interaction processes due to the long lifetime of the mother particles \cite{bib:gaisser}.  Above these energies, the contribution of short-lived hadrons becomes dominant because these follow a harder energy spectrum, similar to that of the primary particle since their livetime is way too short to allow any of them to interact before decaying in the atmosphere, thus always producing neutrinos.

\begin{figure}
\centering
%\begin{subfigure}[horizontal][width=\textwidth]
\includegraphics[width=0.6\textwidth]{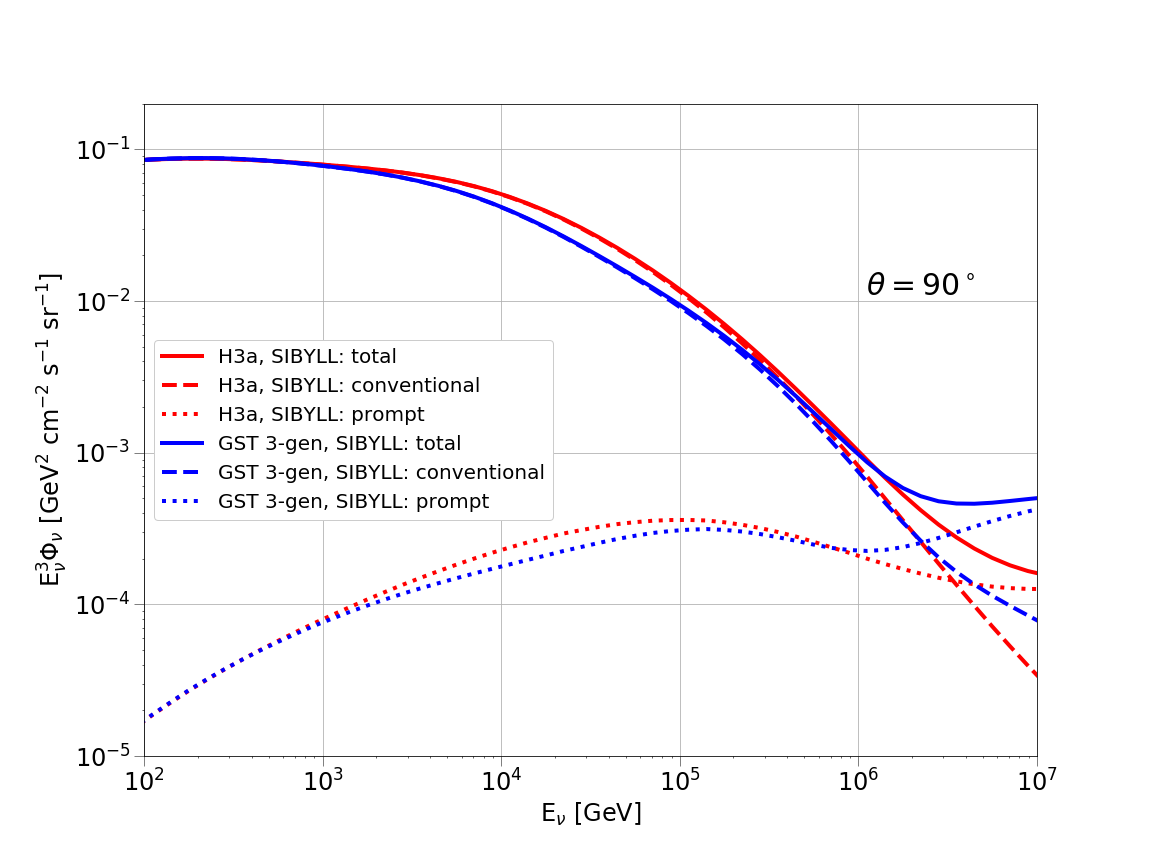}
%\end{subfigure}
%\begin{subfigure}[intermediate][width=\textwidth]
\includegraphics[width=0.6\textwidth]{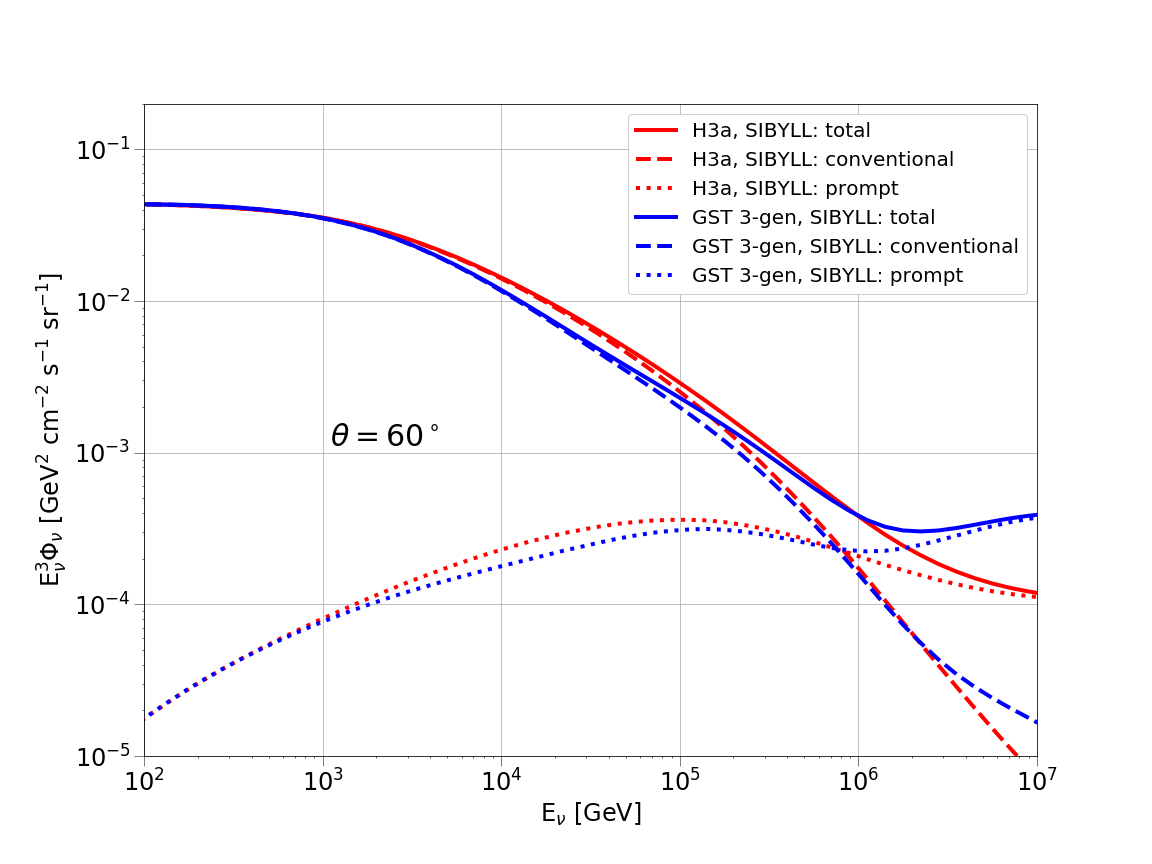}
%\end{subfigure}
%\begin{subfigure}[vertical][width=\textwidth]
\includegraphics[width=0.6\textwidth]{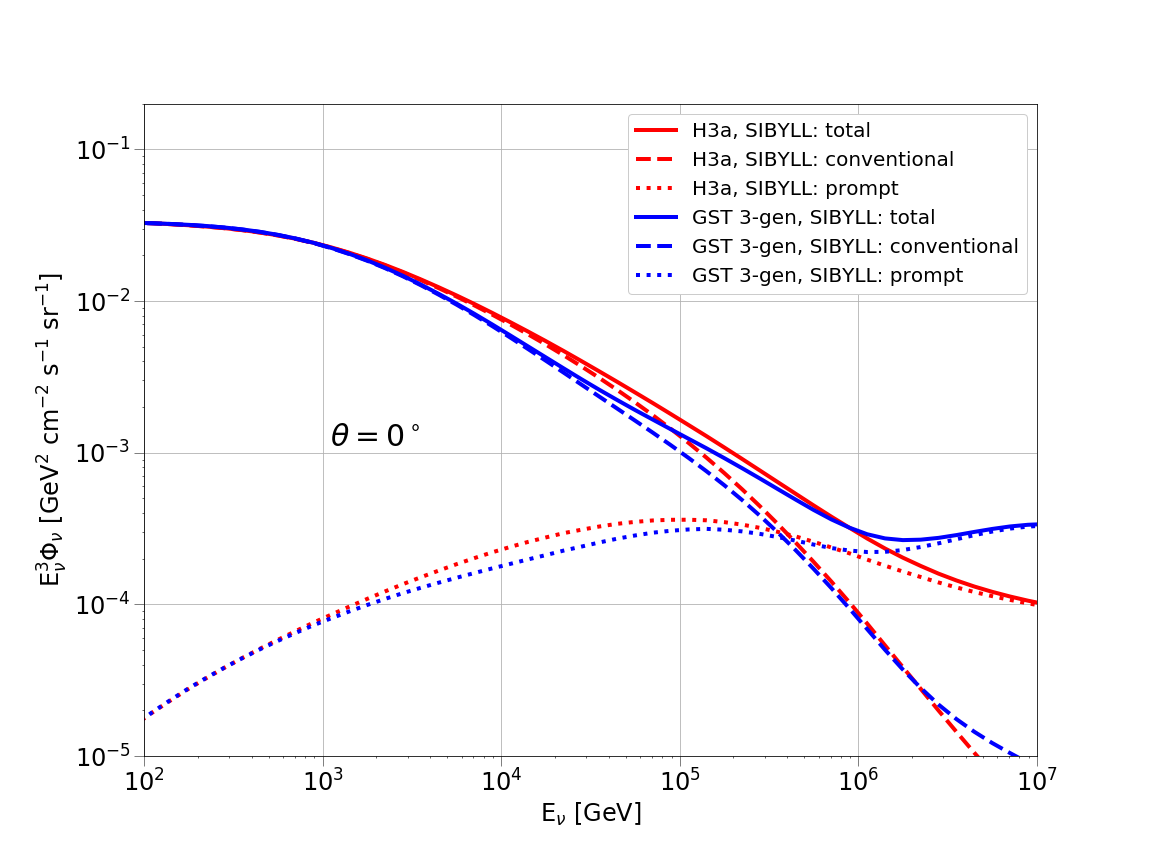}
%\end{subfigure}
\caption{Conventional (dashed) and prompt (dotted) components of the atmospheric $\nu_\mu +\bar{\nu}_\mu$ flux as a function of the energy, multiplied by E$^3$. The red lines are for the flux computed using the H3a CR composition fit, while the blue lines are for the GST 3-gen fit, both considering the SIBYLL-2.3c interaction model. Three zenith angles are chosen: horizontal neutrinos (top), intermediate angles (middle) and vertical (bottom). The solid lines represent the total flux.}
\label{fig:flux_1D}
\end{figure}

In figure \ref{fig:flux_1D} the expected muon neutrino flux for the conventional and prompt component is show for events from zenith equal to 90$^\circ$, 60$^\circ$ and 0$^\circ$ according to the H3a and GST 3-gen CR composition fits. A mid-latitude atmosphere is considered. Two effects are visible: since the prompt flux is constant over zenith while the conventional flux is largely zenith dependent, the influence of the prompt component is much more evident for vertical events than close to the horizon. A large difference between the two CR parameterisation appears in the 10-100 TeV energy range and above some tens of PeV; however, only the former would be visible in the current generation of neutrino telescopes since the neutrino detection rate above the PeV threshold is very low and dominated by the diffuse cosmic signal that, if isotropic as currently claimed and if extending up to those energies, would make any detector blind to this feature.

\begin{figure}

%\begin{minipage}{.5\linewidth}
\centering
%\subfloat[CR composition fit]
\includegraphics[width=\textwidth]{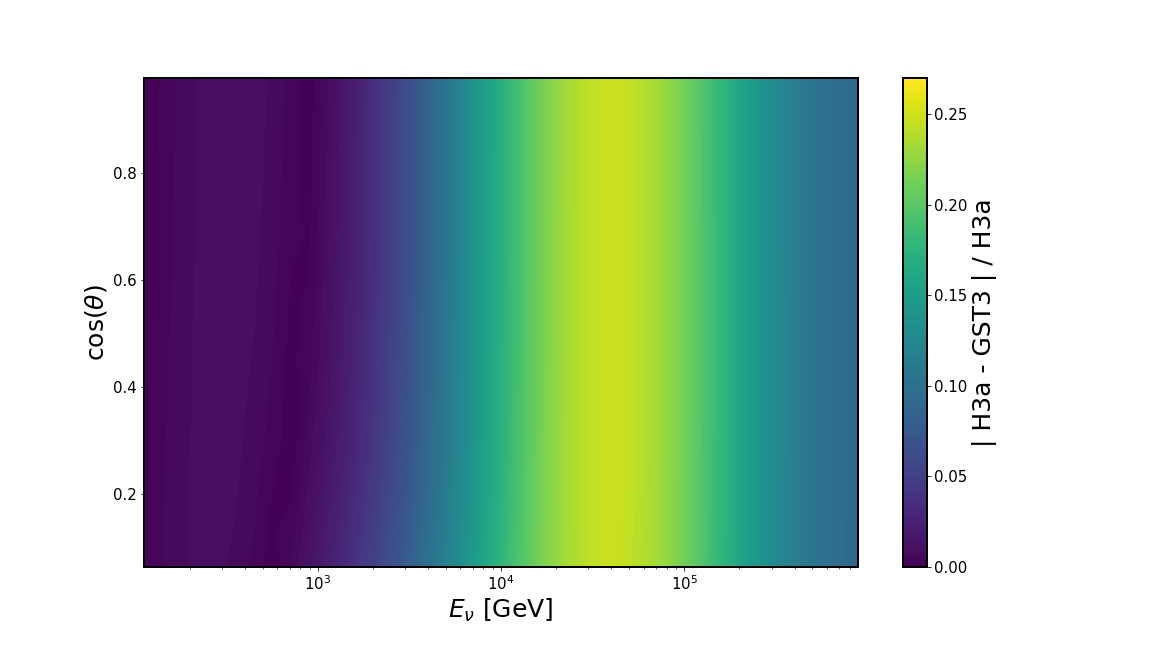}
\includegraphics[width=\textwidth]{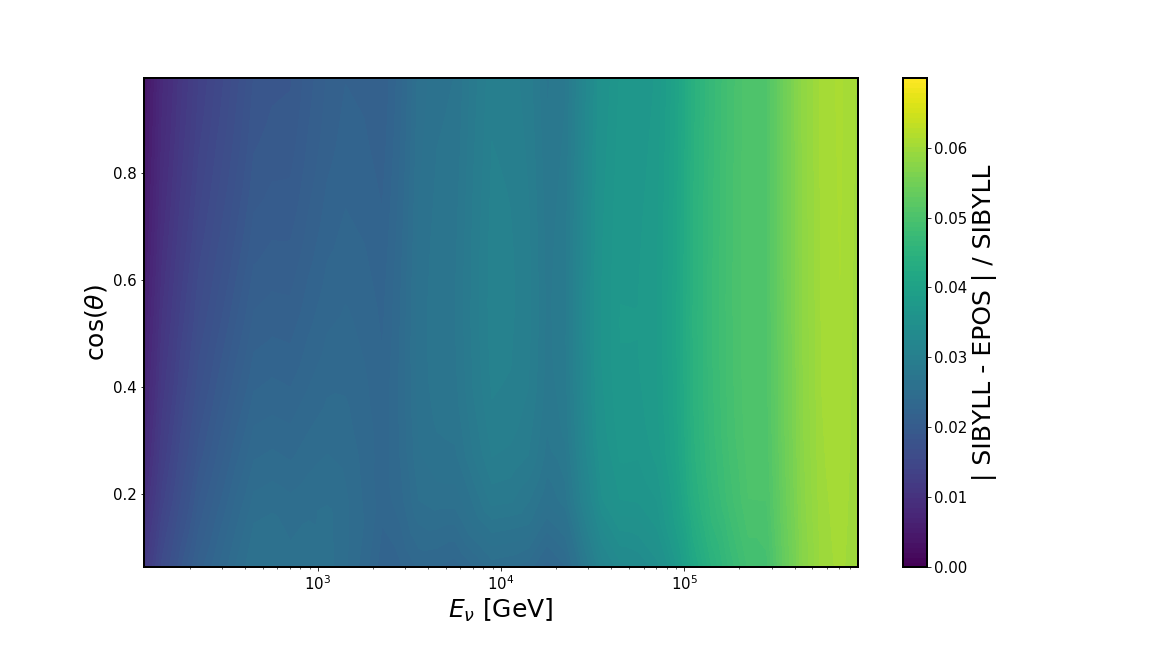}
%\end{minipage}
%\begin{minipage}{.5\linewidth}
%\centering
%\subfloat[CR interaction model]\includegraphics[width=\textwidth]{plots/SIBtoEPOS_ratio.pdf}
%\end{minipage}
\caption{Top: 2D distribution of the relative difference, in absolute value, between the $\nu_\mu +\bar{\nu}_\mu$ conventional fluxes produced by the H3a CR composition fit against the GST 3-gen. Bottom: same for the conventional fluxes produced by the H3a CR composition fit when considering the SIBYLL 2.3c interaction model against the EPOS-LHC model. The 2D distribution are produced in the true neutrino direction, $\theta$, and energy, E$_\nu$. Please note the different range in the color scale.}
\label{fig:2d_fluxes}
\end{figure}

In figure \ref{fig:2d_fluxes}-left the zenith-dependent relative difference between the conventional muon neutrino flux predicted when assuming the H3a CR composition fit versus the GST 3-gen is show. The relative difference also appears to be slightly zenith dependent and, since the neutrino zenith is properly reconstructed in neutrino telescopes, such features would still be present when considering the effect of the detector response. On the right of figure \ref{fig:2d_fluxes} the same kind of plot is reported, but now taking into account the differences induced when considering a different CR interaction model (in particular SIBYLL 2.3c compared to EPOS-LHC). In this case the overall effect is less significant, being of the order of few percent, against the difference coming from the CR fit which is of the order of 10-20\%. Its influence will be considered sub-dominant here and an estimation of the contribution of this effect will be provided as a systematic uncertainty. It should be noted that in both plots the differences above 100 TeV would be smeared out by the existence of the cosmic neutrino flux, which would behave in the same way regardless of the neutrino flux model and thus produce null differences. Its influence will be discussed in the next section when considering the simulated analysis at a VLV$\nu$T.

An additional factor that could be considered when comparing predictions of atmospheric leptons is the nature of atmosphere in which the CR interaction takes place. Differences in temperature directly translate into differences in the air density and thus in the CR interaction target density, as well as in the density of the medium traversed by hadrons and leptons along the atmosphere column. In particular, in the case of polar atmospheres the differences between summer and winter can be remarkable and thus the zenith-dependent flux can be largely affected. This effect is expected to be milder at intermediate latitudes since the summer/winter asymmetry is less pronounced. These differences are significant in the IceCube analyses but will not be considered any further in this particular study. It should be however noticed that this is an effect that must be taken into account when considering a higher-level analysis with real data, which goes beyond the scope of this paper.

%% file: stat.tex
\section{A Monte Carlo study of atmospheric neutrino fluxes} \label{sec:anal}

\subsection{Ingredients for the analysis} \label{sec:ingredients}

The neutrino event rate at the detector induced by a certain neutrino flux is determined by the experiment effective area. The neutrino effective area is dependent on the neutrino interaction cross section, both for what concerns the neutrino interaction rate in the proximity of the instrumented volume and neutrino absorption through the Earth, on the matter density surrounding the detector, on the neutrino detection efficiency of the apparatus, and on the event selection efficiency of the analysis. Because of these factors, the effective area depends on both the neutrino energy and zenith. Neutrino telescopes are usually assumed to be homogeneously sensitive in azimuth. Being the atmospheric neutrino flux (at least for the conventional component) also zenith dependent, the event distribution at the detector must be computed considering both the energy and zenith angle of the incoming neutrino.

\begin{figure}
  \centering
  \includegraphics[width = \textwidth]{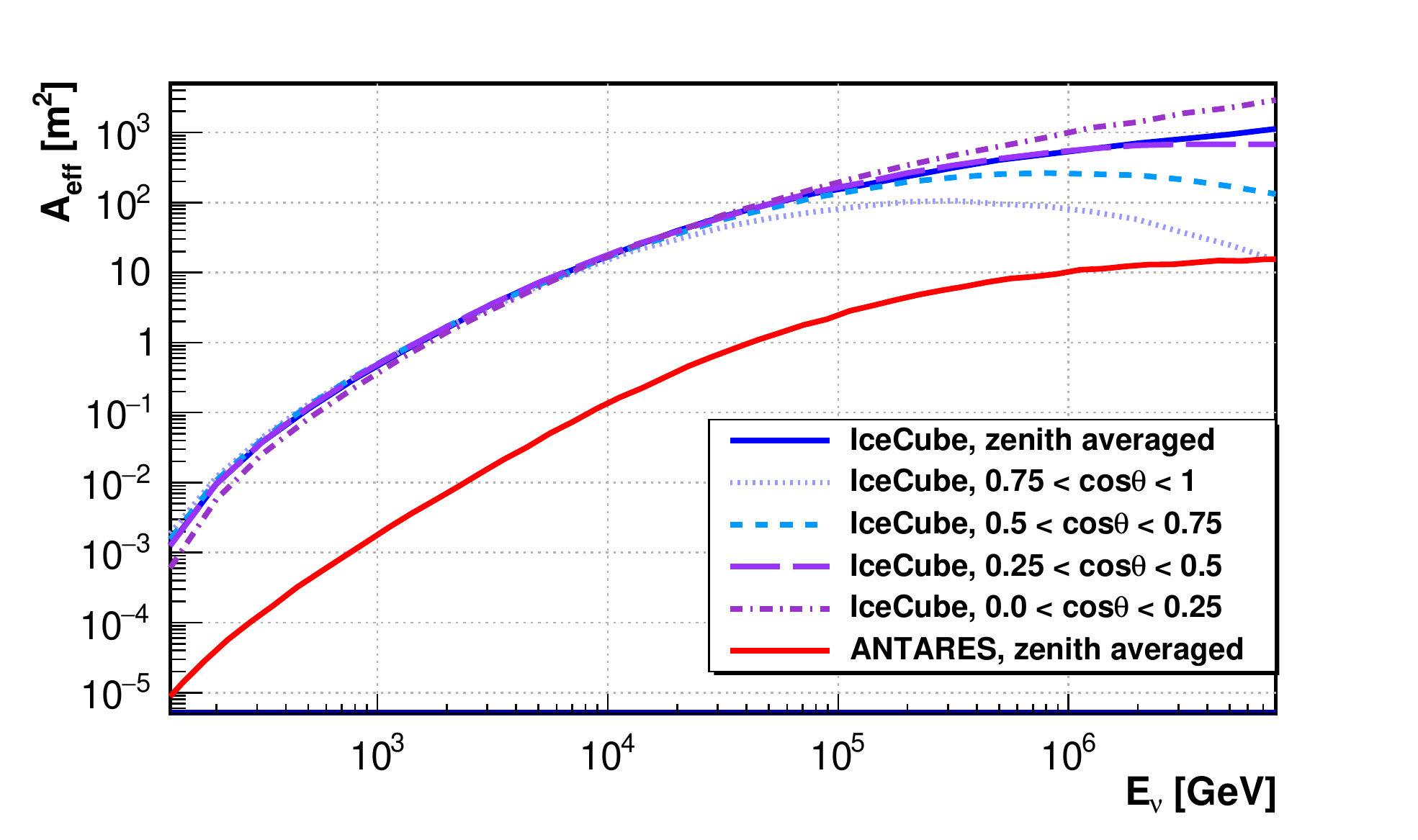}
  \caption{The average between $\nu_\mu$ and $\bar{\nu}_\mu$ effective area of the ANTARES \cite{bib:ant_dif} and IceCube \cite{bib:ic_icrc19_numu} neutrino telescopes are shown as a function of the neutrino energy as solid lines. In the case of the IceCube detector also the zenith dependent ones are shown, as dashed lines, in four bins of the neutrino zenith angle.}
  \label{fig:aeff_all}
  \end{figure}

The IceCube effective area used in this work has been obtained from reference \cite{bib:ic_icrc19_numu}, already provided in bins of zenith angle. This area corresponds to the one after the final event selection, and can thus be directly used to compute the detected event rate. For ANTARES, a zenith-averaged effective area was used in the search for a diffuse flux of cosmic neutrinos \cite{bib:ant_dif}. We obtain the corresponding effective area in different zenith ranges assuming that the selection efficiency is negligibly dependent on the neutrino zenith and thus the zenith dependency is only due to neutrino absorption through the Earth, which can be approximately estimated. Even though this approximation is not really correct, it is sufficient for the scope of this paper.% Similarly, the KM3NeT/ARCA effective area from \cite{bib:km3_loi} is a zenith-averaged effective area and the same correction should be applied to obtain the zenith dependency. In addition, that effective area is reported at trigger level, while the one after the final event selection is required here; in any case, the selection efficiency for the diffuse flux analysis is also reported in \cite{bib:km3_loi} and this efficiency correction is thus applied. 
The zenith-averaged effective areas, plotted as the averaged between $\nu_\mu$ and $\bar{\nu}_\mu$, are shown in figure \ref{fig:aeff_all}. In addition, the zenith dependence of the IceCube one is also presented, to show the order of magnitude of the effect. %.It should be noted that both the IceCube and ANTARES effective areas come from recently published analysis searching for atmospheric and cosmic neutrinos over a large energy range, whereas the KM3NeT/ARCA can be probably considered non-optimal, especially in the 1 -- 100 TeV energy range since that analysis was explicitly searching for very high energy neutrinos only.
The effective areas from KM3NeT/ARCA and GVD-Baikal for an atmospheric neutrino-dominated sample have not been released and thus will not be used in this work. It should be noted that both experiment are expected to reach a size similar to that of IceCube within the next decade, with comparable effective areas.

The event rates at the detector as a function of the true neutrino energy are obtained by multiplying the zenith-dependent atmospheric muon neutrino flux (as in section \ref{sec:crnu}) by the effective area described here above, taking into account the observed solid angle and by the observation time. Since these distributions are obtained in the true neutrino energy while in VLV$\nu$Ts, the neutrino energy is not directly measured for $\nu_\mu$ events a smearing of these distributions is required. Indeed, for almost all events, the resulting muon track is only partially observed. This muon may have travelled several kilometres before being detected and thus only part of its original energy is observable at the detector. Analogously, part of the neutrino energy is released in an hadronic shower which is not directly observable since it might also be kilometres away from the apparatus. Finally, the energy of the muon at the detector is affected by a some uncertainty since the energy estimation in VLV$\nu$Ts is based on sampling energy losses within the instrumented volume -- a quantity that is proportional to the muon energy above a TeV -- which is a stochastic process. In order to obtain a model of these effects, the event rate distributions are smeared so that these energy-estimation-related uncertainties can be accounted for. Since the correct and complete treatment of these effects would require full Monte Carlo simulations of the detector, a simplified smearing is applied here; in particular this smearing corresponds to a Gaussian distribution with width equal to the reported energy resolution of ANTARES \cite{bib:dedx_ant} and IceCube \cite{bib:dedx_ic}.% and ARCA \cite{bib:jgandalf} individually.
This allows us to tentatively reproduce what the detected and estimated neutrino energy distributions should look like in a VLV$\nu$T.

In addition to the smearing effect due to the limited energy resolution, an energy scale uncertainty can affect the measurement. This effect corresponds to a systematic shift of the muon energy estimation, which could also be energy dependent (and with a non-linear dependence). This effect could be induced either by a limited knowledge of the light propagation properties in the medium or of the optical module efficiency, or by a combination of these effects. This systematic shift can be considered as an \textit{unknown unknown, i.e.} a systematic effect that could affect data collected by a VLV$\nu$T and might not be present in Monte Carlo simulations, e.g. because of the lack of an energy calibration. Indeed, the most reliable energy calibration source in a VLV$\nu$T is the flux of atmospheric leptons itself and this energy scale effect can be in principle fitted under a certain number of assumptions when doing the measurement of the energy spectrum. Still, its limited knowledge can strongly affect any energy-dependent measurement such as that of the atmospheric or cosmic neutrino spectrum. Results will be presented here assuming that the energy smearing might either not have such a bias or it might be affected by it, thus the Gaussian smearing would have a certain width, given by the energy resolution, but also it might not be centred at zero.

\begin{figure}
    \centering
    \includegraphics[width=\textwidth]{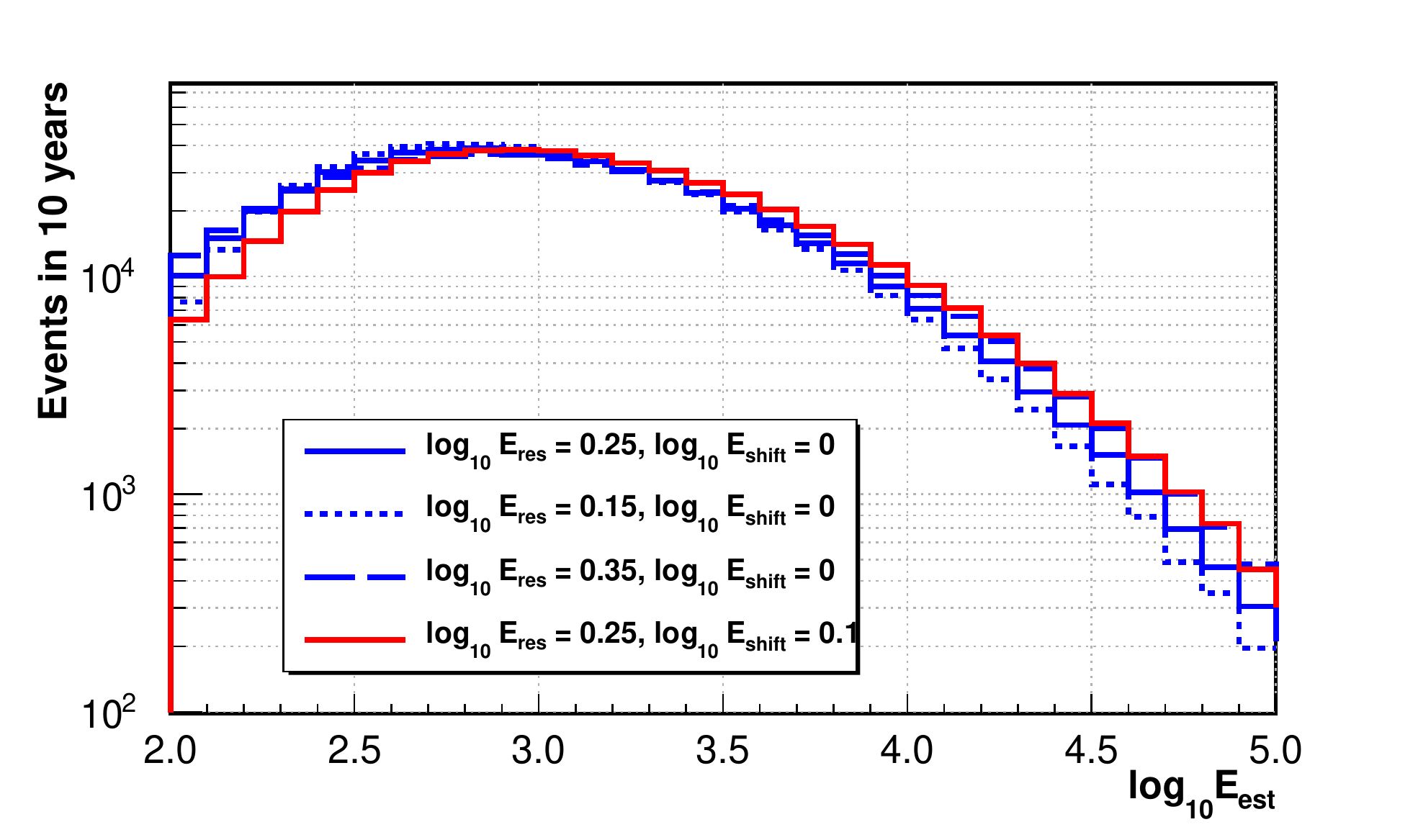}
    \caption{Effect of the energy resolution smearing (blue lines) as well as of the energy scale shift (red line) in the observed event rates in an individual pseudo-experiment reproducing 10 years of zenith-integrated data taking of a detector with the effective area of an IceCube-sized neutrino telescope, for the H3a CR fit.}
    \label{fig:ic_res_rates}
\end{figure}

In figure \ref{fig:ic_res_rates} an example of the energy estimator distribution is shown for different values of energy resolution and energy shift, after the simulation of 10 years of data taking of an IceCube-like detector.
Having defined the log-energy resolution as
\begin{equation}
E_{res}  = log_{10} \frac{E_{reco}}{E_{true}}
\end{equation}
three values are considered, with $E_{res} =$ 0.15, 0.25 and 0.35 in the logarithm of the true energy, as well as an energy shift equal to 0.1 in the logarithm under the assumption that the resolution is 0.25. Changing the energy resolution modifies the spectrum making it smearing out differences and allowing more low-energy events to migrate to the high-reconstructed energy bins. This is due to the fact that the atmospheric neutrino energy spectrum is steeply falling and thus more events are expected at low energies with respect to the highest energies. For this reason, the probability that a low-energy event would pollute the high-energy queue of the distribution is higher.

An energy shift would mostly affect the low-energy region, causing a shift in the turn-on regime, \textit{i.e.} in the energy range where the detector effiency is lower; at the highest energies also this effect would be smeared out by the resolution.

\subsection{Statistical analysis tools}  \label{sec:stat}

A binned maximum likelihood approach is followed, similarly to what has been done by the ANTARES and IceCube Collaborations in their searches for a diffuse flux of cosmic neutrinos \cite{bib:ic_comb}\cite{bib:ant_dif}. It should be noted that the results of a search similar to the one presented here has been presented by the IceCube collaboration in reference \cite{bib:ic_icrc19_numu}: the scope of this paper is however to understand to which extend such a procedure should be improved for a VLV$\nu$Ts to study neutrino fluxes and how much improvement in the current reconstruction performances and statistics would be needed to constrain different atmospheric neutrino flux models. Most of the statistical tools used in this work are taken from the ROOT \cite{bib:root} and RooFit \cite{bib:roofit} frameworks. %similar to the one presented here has been used in the IceCube data analysis.

Pseudo-data sets are generated from different models and the distribution of a simulated energy estimator is produced, accounting for the detector energy resolution smearing and allowing for an unknown energy scale systematic effect to influence the distribution, as described in the previous section. Together with this, also a  normalisation systematic effect is present in this simulated data, allowing the overall flux to fluctuate around the expected value. In this work, this possible fluctuation is assumed constant over the full energy range energy, even though the flux normalisation uncertainty is energy dependent and probably a more complex influence, such as a polynomial behaviour of this uncertainty with energy could be present \cite{bib:bartol_syst}. An energy-uniform effect, as well as a normalisation effect which is linerly-dependentt on the energy -- not so dissimilar to an energy scale effect -- can be fitted by the likelihood procedure without affecting significantly the result. Higher order effects have also been considered in The latest IceCube analysis: the fitting procedure has been shown to be robust against these effects as well \cite{bib:ic_icrc19_numu}.

Similarly, an energy-uniform tilt in the spectral behaviour would induce an overall shift in the data, which is analogous to that assumed to come from a possible energy mis-calibration of the detector. The two effects are not factorisable with the current approach. In addition to the smearing effects due to the neutrino energy reconstruction, the distributions are effected also by statistical fluctuations, considered here as poissonian given the large per-bin statistics expected in a km$^3$-scale VLV$\nu$T.

The zenith- and energy-binned distribution of pseudo-data is compared to the one expected from the model, smeared as well by the expected energy resolution of the experiment, but only by this. The flux normalisation is fitted using a likelihood maximisation algorithm, separately for the conventional, the prompt and the cosmic components since the three are affected by different uncertainties, most of which are uncorrelated. The likelihood function used here, $L (d | h)$, is given by the product of the individual likelihoods $L_{i} (d | h)$ computed for each bin $i$ of the energy distribution observed in pseudo-data sets compared to those of the models. The individual likelihood is given by a Poisson probability term; in particular
\begin{equation}
\displaystyle L (d | h) = \prod_{i=0}^{N} L_{i} (d | h) 
\end{equation}
where, dropping the $(d|h)$ from the formula,
\begin{equation}
    L_{i} = e^{-\mu_{i}}\cdot\frac{{\mu_{i}}^{k_{i}}}{k_{i}!}%\cdot %\prod_j\frac{1}{2\pi\sigma[\tau_{i,j}]}e^{-\frac{\left(\tau_{i,j}-\tau_{i,j}^*\right)^2}{\sigma^2[\tau_{i,j}]}}  
\end{equation}
$\mu_i$ is the expected number of events in the $i$-th bin from the simulated templates according to the hypothesis $h$, $N$ is the number of bins in the energy estimator histogram for each event sample and $k_{i}$ is the number of events observed in pseudo-data $d$ for that event sample in that bin.

\begin{figure}
    \centering
    \includegraphics[width = \textwidth]{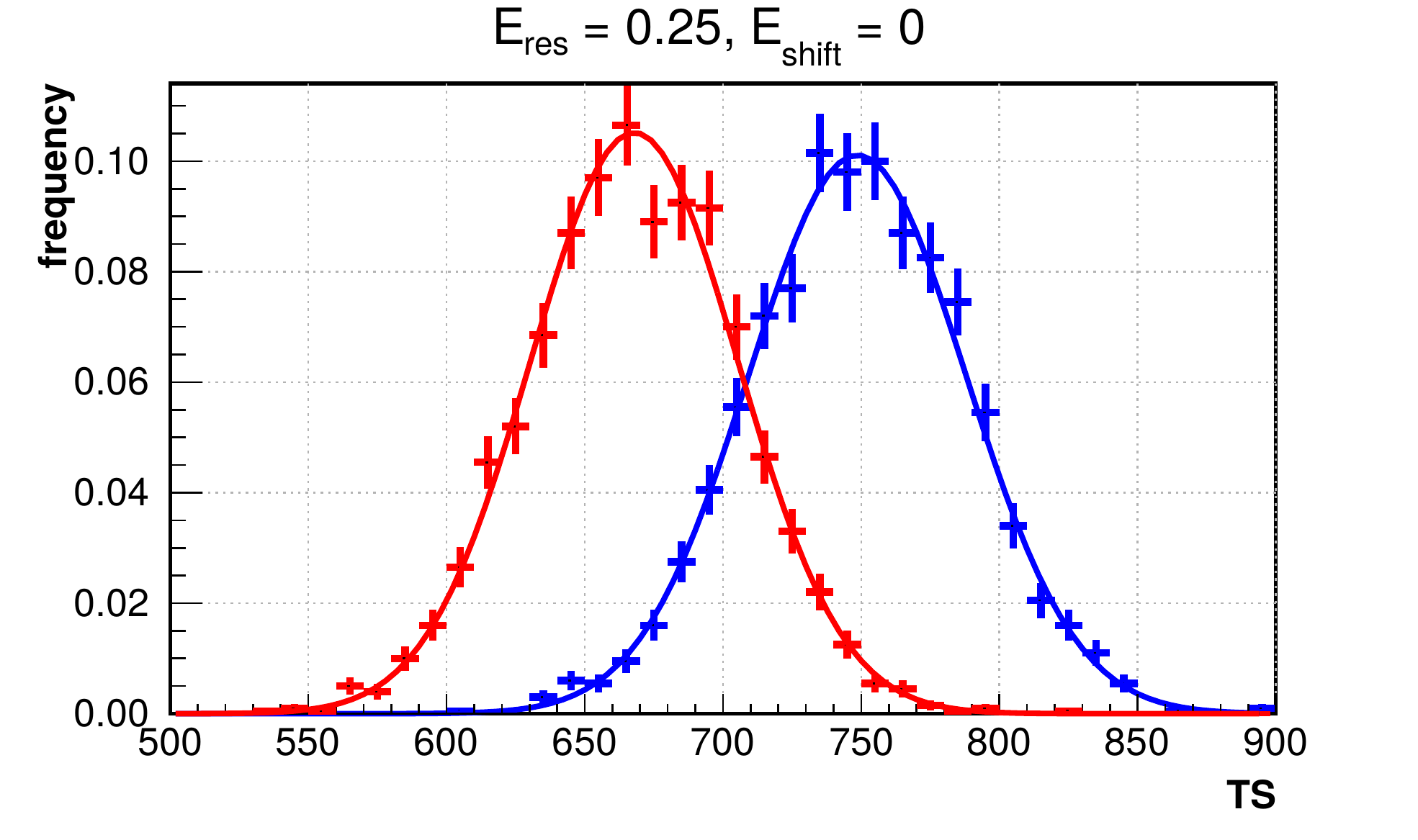}
    \caption{TS distribution (frequency, being the distribution normalised to the total number of performed pseudo-experiments) for the H3a (red) and GST3 (blue) in a particular combination of energy shift and resolution, coming from 2000 pseudo-experiments, each corresponding to 10 years of an IceCube-sized neutrino telescope data taking.}
    \label{fig:TS_def}
\end{figure}

In order to estimate the sensitivity of an experimental configuration, a test-statistics TS is defined as a log-likelihood difference:
\begin{equation}
    TS = \log L( d_i | h_i) - \log L (d_i | h_j)
\end{equation}
\textit{i.e.} testing the data against one model or the other, and by comparing the difference of the TS distributions obtained for several pseudo-experiments for data generated according to different hypotheses. The TS is distributed as a Gaussian distribution. The \textit{median sensitivity} is defined by considering the area of the Gaussian of the TS distribution beyond the median of the TS distribution obtained by exchanging hypotheses. Figure \ref{fig:TS_def} shows the TS distribution obtained by generating 2000 pseudo-data sets of an IceCube-sized detector with an energy resolution equal to 0.25 in the logarithm of the neutrino energy (and under the assumption of a null-cosmic flux). These pseudo-experiments are generated according to the H3a model (in red) or the GST 3-gen (in blue) and the TS value is computed for each of them. A Gaussian fit is performed on both distributions. Given the mean values of the two Gaussian curves and their variances, the median sensitivity is computed. In this particular, unrealistic case, it corresponds to a p-value of 0.02 meaning that 10 years of an IceCube-sized neutrino telescope with this precision in the reconstruction of the muon neutrino energy would be able to separate the two models to roughly a 2$\sigma$ level. In section \ref{sec:results} several possible outcomes of the procedure will be discussed.

In order to evaluate the impact induced by systematic effects, these could be added as nuisance parameters in the fitting procedure, as done in references \cite{bib:ant_dif}\cite{bib:ic_icrc19_numu}. For simplicity, since not all the detector-related uncertainties can be accounted for here, we computed the effect of systematic uncertainties by means of a $\chi^2$ comparison between the expected distributions at the detector and taking the $\sqrt{\chi^2}$ difference induced by a certain effect as an estimate of the corresponding decrease in terms of significance. The $\chi^2$ is defined as

\begin{equation}
    \chi^2 (H_0, H_1)= \sum_i^{N_{bins}} \frac{(\mu_i^{H_1} - \nu_i^{H_0})^2}{\nu_i^{H_0}}
\end{equation}
where $\mu_i^{H_1}$ is the observed number of events in the i-th bin according to the hypothesis $H_1$ and $\nu_i^{H_0}$ is the expectations from $H_0$. $H_0$ and $H_1$ are modified to account for a systematic effect producing a certain $H'_0$ and $H'_1$ and $\chi^2 (H'_0, H'_1)$ is computed. The influence of a systematic effect is estimated to be of the order of
\begin{equation}
    \Delta \chi ^2 = | \chi^2 (H'_0, H'_1) - \chi^2 (H_0, H_1) |
\end{equation}
being the square root of this $\Delta \chi ^2$ and estimation of the significance drop.

%% file: results.tex
\section{Results} \label{sec:results}

We first analysed the case of a detector with the size of ANTARES. In this case, given the limited effective area of the neutrino telescope, the possible separation between any different model remains at the level of 0.8 in p-value even assuming an extremely optimistic energy resolution of the detector of 0.35 in the logarithm of the neutrino energy. When using a realistic value as from \cite{bib:dedx_ant}, the TS distributions become basically overlapping. The only effect that appears to be measurable within 1$\sigma$ with ANTARES is a change of slope in energy spectrum larger that $| \Delta \gamma_p | = 0.1$. The small ANTARES effective area does not allow the model-induced fluctuations shown in figure \ref{fig:2d_fluxes} to become relevant above the statistical fluctuations. This result is close to what had been reported in reference \cite{bib:ant_icrc19_diff}.

The configurations considered in this work for 10 years of data acquisition of an IceCube-sized neutrino telescope provide some more interesting results. When considering the best fit result from the search of a cosmic flux as in reference \cite{bib:ic_icrc19_numu} and a realistic assumption of the neutrino energy resolution of 0.5 in the logarithm of the energy, a discrimination between the H3a and GST 3-gen model is of the order of 1.8$\sigma$.

The influence of the cosmic flux in this evaluation is estimated by considering the strategy described previously. Different cosmic components are introduced in the pseudo-experiments ranging from the lowest-normalisation and softest-spectrum cosmic fit allowed within 1$\sigma$ in reference \cite{bib:ic_icrc19_numu} to the highest-normalisation and hardest-spectrum one provided in the same reference. As a result, we find that the discrimination power changes by roughly $\pm$0.2$\sigma$ when considering different cosmic spectra. Indeed, for lower cosmic normalisation the discrimination power increases; a similar effect is present for harder cosmic spectra. Prompt neutrino fluxes are currently affected by very large uncertainties. Their contribution can be considered sub-dominant, even though still relevant, because of the presence of a clear cosmic flux which appears in the same energy region. Some indication of its relevance in a full-spectrum analysis can be found in reference \cite{bib:mascaretti}. Dedicated studies should be performed to better understand the relevance of the prompt component in the measured high-energy neutrino fluxes; some discrimination power probably would come from the measurement of electron neutrino spectra in VLV$\nu$T. This channel has not been treated here.

\begin{figure}

%\begin{minipage}{.5\linewidth}
\centering
%\subfloat[CR composition fit]
\includegraphics[width=\textwidth]{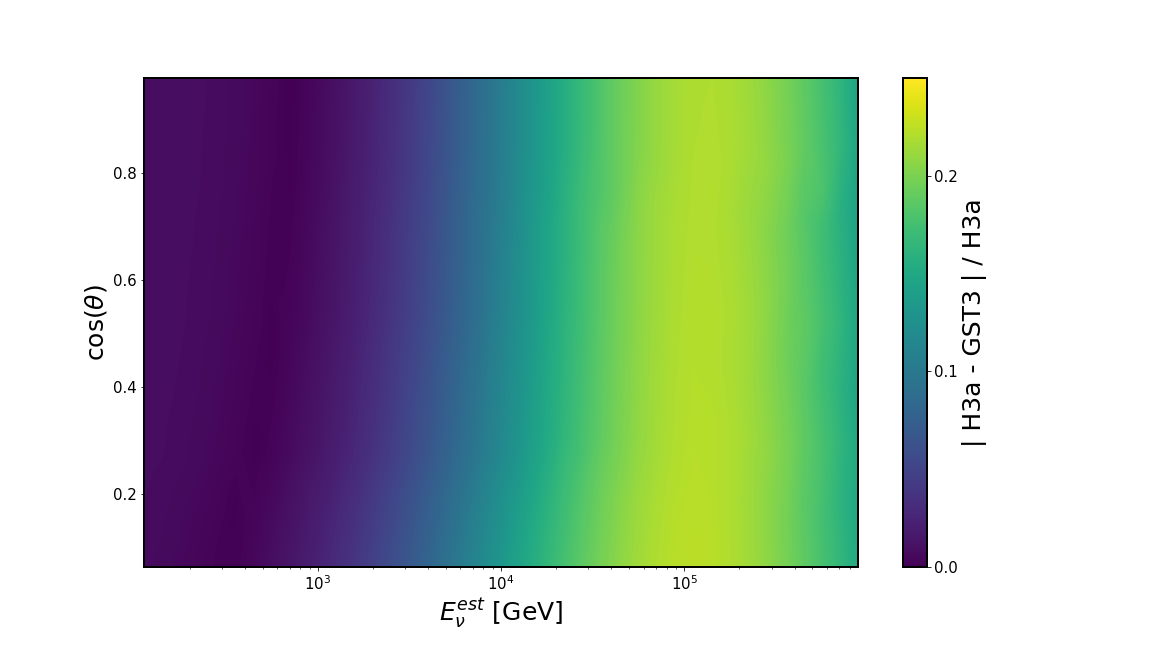}
\includegraphics[width=\textwidth]{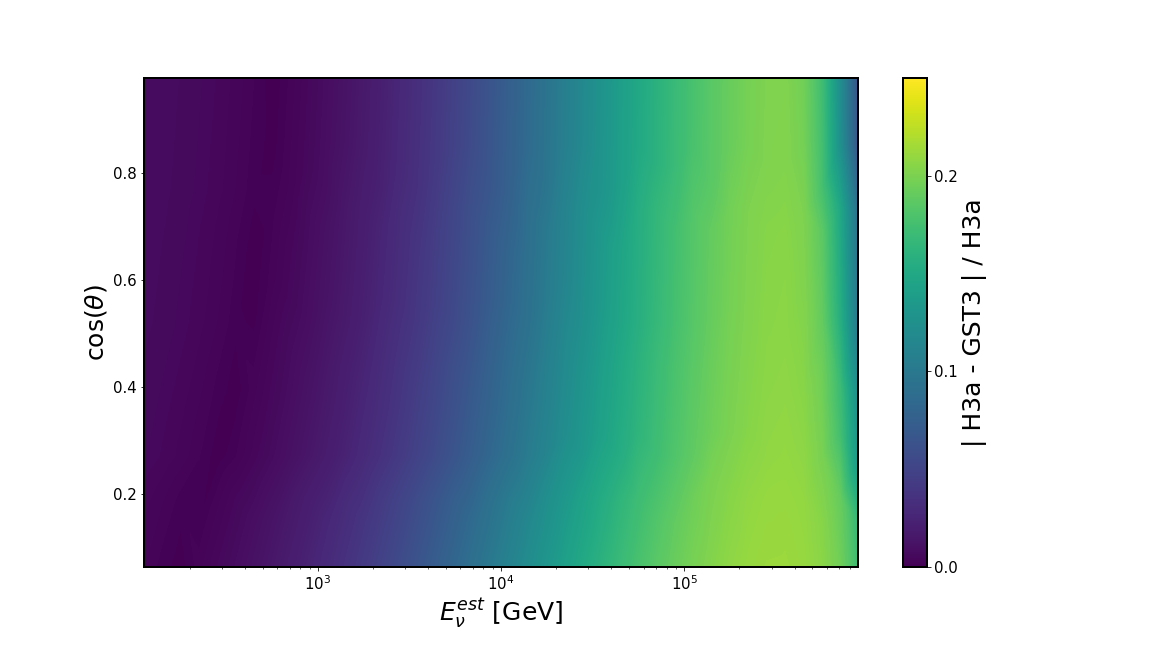}
\caption{2D distributions of the relative difference, in absolute value, between the overall flux expected in 10 years of data acquisition of an IceCube-sized detector including the conventional and prompt component, as well as a cosmic flux equivalent to the best fit proposed by IceCube in reference \cite{bib:ic_icrc19_numu} when considering the H3a and the GST 3-gen model. The top plot is assuming an energy resolution of 0.35 in the logarithm of the neutrino energy, while the bottom plot is for an assumed resolution equal to 0.50. The 2D distribution are produced in the estimated neutrino direction, $\theta$, and estimated energy, E$_\nu^{est}$. }
\label{fig:2d_fluxes_c}
\end{figure}

In order to test the possible limits of this analysis, we have considered an improved neutrino energy reconstruction pushing the IceCube one to the best expected results that could be obtained, e.g., in an underwater neutrino telescope such as KM3NeT/ARCA \cite{bib:jgandalf}. The observed effect corresponds to an improvement of about 0.3$\sigma$ in sensitivity. An improved energy resolution also makes the analysis less sensitive to the effects induced by the uncertainty on the background given by the diffuse flux of cosmic neutrinos. This is partly shown in figure \ref{fig:2d_fluxes_c} where, in the same fashion as in figure \ref{fig:2d_fluxes}, the H3a and GST 3-gen models are compared, once the cosmic background is also added to the observed distributions. However, the effect of the energy smearing is accounted for in this figure, considering a resolution of 0.35 (top) and 0.5 (bottom) in the logarithm of the neutrino energy and adding the expectation from a cosmic flux to that of the atmospheric signal. The presence of a cosmic component removes completely the differences at PeV energies. In a similar way, the limited energy resolution smears the effect because low-energy events, being more abundant, can easily pollute the region where the highest-energy ones are expected.

\begin{figure}

%\begin{minipage}{.5\linewidth}
\centering
%\subfloat[CR composition fit]
\includegraphics[width=\textwidth]{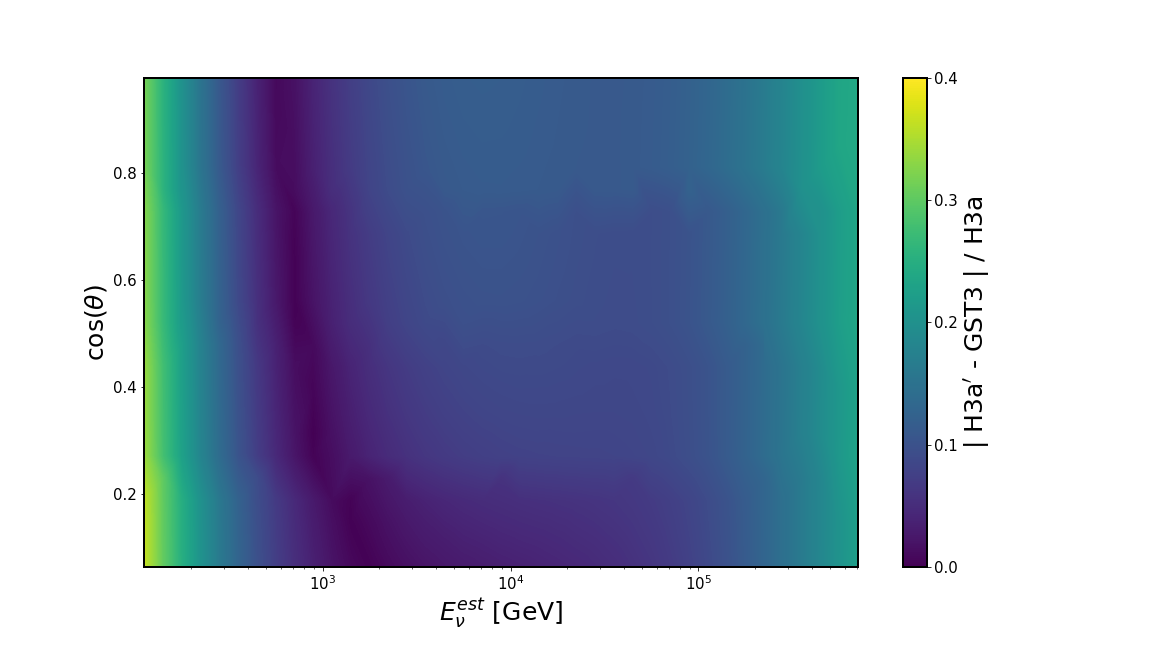}
\includegraphics[width=\textwidth]{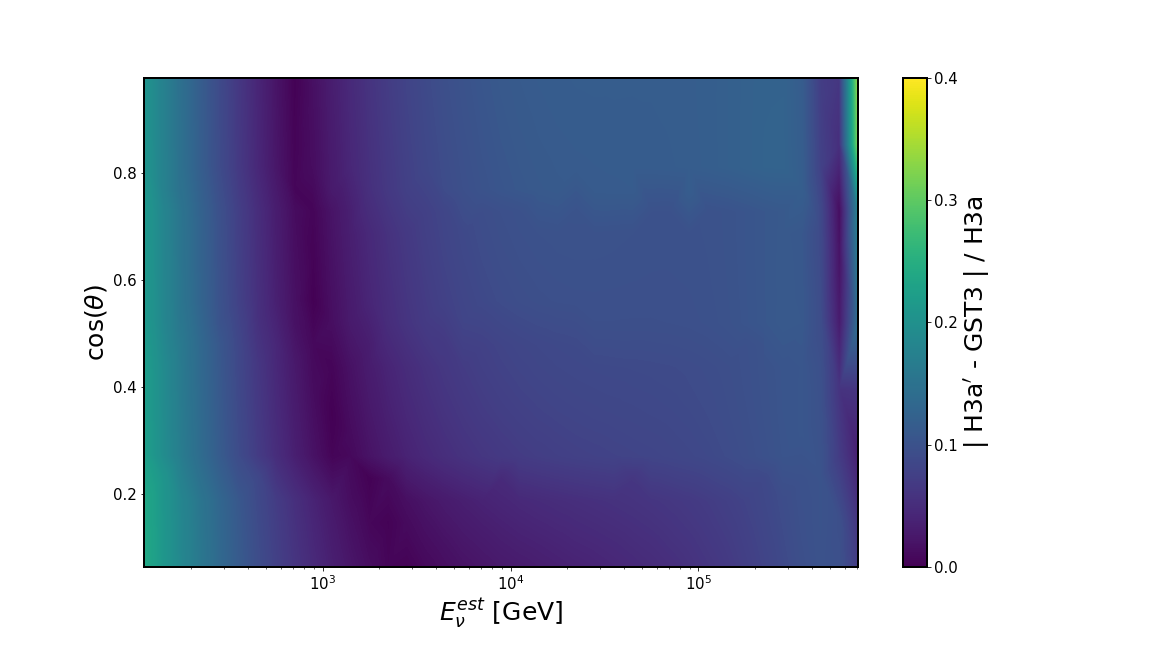}
\caption{Same as figure \ref{fig:2d_fluxes_c} but an energy shift is applied to the pseudo-data from the H3a model (H3a'), which are then compared to the expectation from the GST 3-gen against the H3a without any shift. The top and bottom plot use the same assumption as in the top and bottom of figure \ref{fig:2d_fluxes_c}.}
\label{fig:2d_fluxes_c_sh}
\end{figure}

Finally, the effect shown in figure \ref{fig:flux_1D} given by a possible energy scale uncertainty, has been considered. In order to do so, the standard configuration, e.g. the one corresponding to two plots from figure \ref{fig:2d_fluxes_c}, without any systematic shift has been compared to data-sets where a shift in the logarithm of the energy equal to -0.1 had been applied in pseudo data-sets that were to be compared to models where this shift had not been applied. As a result, as shown inf figure \ref{fig:2d_fluxes_c_sh} the differences between the expectations decrease in the region where it was peaking before by a large factor, and a relevant difference between the distribution appears at the lowest energies while in the standard scenario no difference was visible below 1 TeV. An opposite effect is present when a positive shift is applied. Thus this effect can definitely bias an analysis such as the one described here. This effect is somewhat milder when the energy resolution is improved.

This kind of uncertainty can be accounted for in a full-likelihood fitting procedure as a nuisance parameter, as e.g. done in reference \cite{bib:ant_dif}. If this effect is constant with energy, it can be disentangled from the spectral measurement by using constraints from events below 1 TeV, where the relative differences between models are smaller, the neutrino statistics are larger and the effects induced by energy scale uncertainties are more significant. Given this, it is possible that in a complete analysis this effect might become even milder.

In these computations, the uncertainty given by the direction reconstruction has been neglected; this is surely not affecting the result above 10 TeV, where the angular resolution of VLV$\nu$T is well below the size of the zenith bins considered here. At lower energies, also the angular resolution should start playing a role; however, the contribution to the total significance coming from these lower energies is not too large, and the total effect is only partly zenith dependent in that energy range.

As already mentioned along the paper, some effects have been neglected here and would certainly play a role in the overall estimation of the sensitivity. One effect that could probably be significant when approaching the problem as done here, would be that induced by seasonal variation of the atmospheric lepton fluxes measured at the South Pole; however, the current knowledge of the temperature profile of the atmosphere would probably guarantee that this effect could be kept under-control in a fully defined analysis using complete detector responses.

One further aspects that could induce systematic deviations in the observable zenith-energy distribution at the detector would be related to the limited knowledge of neutrino cross sections at very high energies. Indeed, in these conditions, only VLV$\nu$T can attempt a measurement of this quantity just using atmospheric neutrinos, actually by measuring deviations in the 2D zenith-energy distribution in a very similar way to what has been proposed here \cite{bib:xsec-nu}. If one assumes that the neutrino cross section should follow a scaling by a constant factor (or by a linearly-energy-dependent one) with respect to the standard model prediction, this can be probably absorbed in a full-likelihood analysis as nuisance parameter under the assumption that the inner-Earth model, in terms of density, is known with a much smaller uncertainty. In general, the effect due to a different neutrino cross-section would mostly affect the measurements in the 10-100 TeV range, in the region when absorption through the Earth becomes non-neglibile, where also this analysis is most sensitive. However, effect which depend on the neutrino cross-section would be a more remarkable zenith-dependent behaviour, especially in the region of the Earth core.

Another systematic uncertainty related to the particle physics at play is the one induced by the hadronic model in the CR interaction and shower development in the atmosphere. As shown in figure \ref{fig:2d_fluxes} its influce does not show a huge energy dependent effect and seems quite linear over the energy range. We assumed here that its impact would not be too large when the central expectation values are taken into account. Given the uncertainties on these prediction, we cannot exclude they could significantly change the picture, but probably further improvements on their precision could be foreseen. 

As a matter of fact, a polished, even though model-dependent, analysis could be already performed with the current generation of km$^3$ neutrino telescopes, and the separation power between different models could be of the order of 1.5 to 2$\sigma$; larger-volume neutrino telescopes are foreseen in the next future (KM3NeT/ARCA and GVD-Baikal) or on a slightly longer term (such as IceCube-Gen2 \cite{bib:gen2}). Given the improvement in performance expected for these instruments, a better insight on this question could be given if large and pure atmospheric neutrino data-sets are collected, benefiting from both the larger exposure and the more precise measurement of the muon neutrino energy. The possibility of combining different data-samples collected by different telescopes should not be underestimated, since this would help controlling detector systematical uncertainties e.g. in the energy calibration which could be of great importance in the proper estimation of the behaviour of atmospheric neutrinos. Finally, expanding this to the electron neutrino channel might be challenging because of the lower statistics achievable and the necessity of considering further uncertainties coming from neutral current events, but this channel might also be worth testing in the future.

\section*{Acknowledgements}

\footnotesize{The authors would like to thank V\'eronique Van Elewyck and Antoine Kouchner for having read and commented this manuscript.

  L.A. Fusco acknowledges the support of the Agence Nationale de la Recherche (contract ANR-15-CE31-0020). 

\noindent L.A. Fusco is also grateful to the Mainz Institute for Theoretical Physics (MITP) of the DFG Cluster of Excellence PRISMA$^+$ (Project ID 39083149), for its hospitality and its partial support during the completion of this work.}